# Enacting Planets to Understand Occultation Phenomena

**Emmanuel Rollinde[1]**

[1]Laboratoire de Didactique André Revuz, EA 4434, Université Cergy-Pontoise, 33, boulevard du Port, France; emmanuel.rollinde@u-cergy.fr

*Last version before publication*

**Abstract**

The Solar System motivates students to interest themselves in sciences, as a large number of concepts may be easily introduced through the observation and understanding of the planet's motion. Using a large representation of the Solar System at a human scale ("a human Orrery"), we have conducted different activities with 10 to 16 years old children. In this contribution, we discuss the different scientific concepts covered by the Human Orrery, allowing the connection of both science and mathematics subjects in schools. We then detail how this pedagogical tool may serve to introduce abstract concepts required to understand occultation phenomena through a modelling activity.

## 1. Introduction

In 2013, a program called "kinaesthetic activities in teaching science and humanities" was granted by Sorbonne University, France, connecting UPMC (departments of physics and biology), and Paris Sorbonne (departments of sports, Italian, and ancient Greek). A "human Orrery" was thus created by Rollinde et al. (2015), allowing the learners to enact the planets' movement with correct relative speed (Figure 1). An Orrery is a mechanical or digital device designed to illustrate the motion of the planets around the Sun and their changing positions in the sky. On a human Orrery, the orbits of planets and comets are drawn at a human scale allowing movements in the Solar System to be enacted by the learners. Astronomy, in general, provides a highly motivating context for learners to develop observational skills, discover methods of scientific inquiry, and explore some of the fundamental laws of physics and concepts of mathematics in both an attractive and meaningful way. The implementation of Human Orrery, as a learning experience, involves topics from mathematics and science and illustrates an example of a STEM approach of learning about concepts perceived as abstracts by students and teachers.

Through our initiative, seven Human Orreries were built in France (one for a science center, one in a public place in Paris – Le Jardin des Plantes, five in primary or secondary schools), one was drawn in a Lebanon school and one map was purchased by a science center in Vietnam. Human Orreries are used thereafter by different teachers (physics, mathematics, technology) in those places. The topics of different sequences tested since 2015 are described briefly. Details may be found in the references or through direct requirement to the author. (i) Construction of a human Orrery. The description of the Solar System involves different length scales from the diameter of the planets to the distance between planets and the length of one orbit. It also involves different duration scales from the rotation to the orbital period. The construction of a Solar System implies a choice of length scales and orientation and the drawing of ellipses. The links with mathematics are obvious, including placement on a 2D plan, Euclidian division and simple geometry. (ii) Enacting planets to learn physics. In the last 3 years, planets have been enacted by undergraduate students in Paris as well as pupils of primary and secondary schools in Paris, Nantes, and Beirut on their own Orrery or on Sorbonne University one. Through the eccentric orbit of a comet, students have observed its varying velocity. They then realized why force and speed are not directly related, and how the work-energy theorem is connected with real movements… 16-year-old students have observed, measured and plot Kepler's laws and experiment different referential frames in astrophysics context (Rollinde, 2017; Rollinde & Decamp, 2019) with significant improvement in their understanding and use of change of frames. 12-year-old pupils have experienced the complex relationship between duration, distance, and velocity with a specific insight that a larger distance may be travelled in a larger amount of time if the velocity is smaller (Abboud et al. 2019). 10-year-old pupils have made a movie about shooting stars (encounter of Earth and a comet during the night, on planetaire.overblog.com) and pre-service teachers were introduced into the use of the Human Orrery and astronomy, in general, to set up different sequences on geometry, velocity, localization of planets and orbital periods.

In this contribution, the design and operation of the Human Orrery will be shortly described, together with a short theoretical background on enaction theory. Then, a specific proposition that concerns specifically occultation phenomena will be presented.

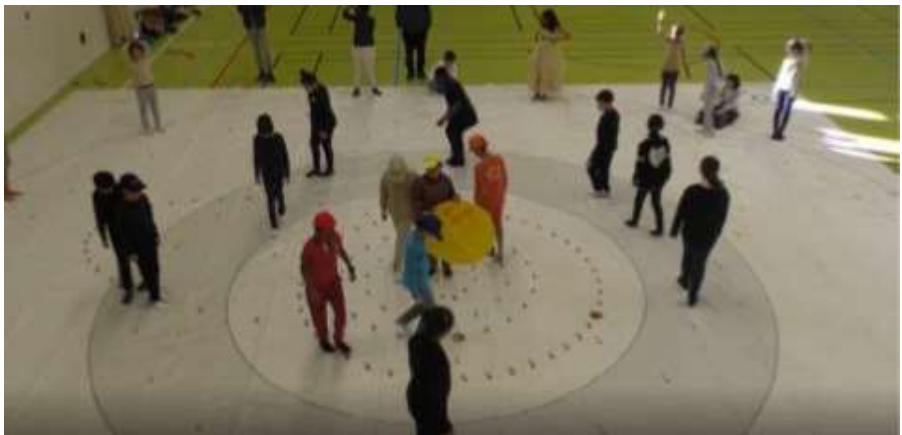

**Figure 1. The "Human Orrery": Pupils enact planets (Mercury to Jupiter), asteroids and comets (in black) along their orbits around the Sun**

## 1. The Human Orrery
### 1.1. A Short Manual
The design of a Human Orrery is made such that users may walk along orbits of different bodies around the Sun that is located at the center of the design. The Human Orrery that we use is printed in a large map of 12m by 12m (Figure 1). It allows one to follow the orbits of the inner planets (Mercury, Venus, Earth, and Mars) and Jupiter; the inner planets are located inside the asteroid belt that is materialized with a grey colour together with the orbit of the largest asteroid known, Cérès. The highly eccentric orbits of two comets are also used: Encke and Chury. This choice of objects illustrates different types of movements while keeping the size of the Orrery reasonable. Earth is at one meter from the Sun (spatial scale), while Jupiter's orbit has a diameter of 10,5m.

The orbits of all bodies are materialized by dots at constant intervals of times, with accurate elliptical shapes. Note that orbits are near-circular for the five planets and Ceres. The interval of time may be different for each orbit, but is always a multiple of 16 terrestrial days: For Earth, there are 23 dots separated by 16 days, which would make a period of 368 days. For Jupiter, there are 54 dots separated by 80 days, which makes a period of 4320 days instead of the real period of 4332,59 days. Note that this difference creates a difference between the walk on the Orrery and the walk of the real planet at each round–about 3 days too late for Earth which combines into 15 days after 5 terrestrial years. To account for this difference, one has to jump one point after 5 round on Earth. This may be used as an introduction to the Euclidian division in mathematics – how to divide 365,25 days into intervals of 16 days (365,25 = 23*16 – 2,75). A sound (either a clock or hand claps) is heard regularly. Every user makes one step during this time interval. The person that enacts Earth walks from one point to the next (distance) in one step (duration). Hence, the interval of time between two sounds corresponds to 16 terrestrial days (temporal scale). "Jupiter" has to do five steps (five times 16 days) to reach the next point. All rules are described in Rollinde (2017) and Rollinde & Decamp (2019). By acting according to those rules, the movements on the Human Orrery illustrate the correct relative velocities of all Solar System bodies.

**1.2. Scientific Concepts in Use**

We review now the concepts associated with the solar system in general and the Human Orrery specifically, and the conceptions that may hinder this knowledge. The order chosen follows the chronology of learning in the French school context.

Firstly, the factual knowledge of what makes up the solar system is part of the astronomy literacy that must be taught at school, including the great diversity of bodies in the solar system, from small rockets (asteroid and comets), small planet with or without atmosphere, telluric planets, gaseous planets... The words "meteor, meteorites, shooting stars" will come into the discussion and allow the teacher to clarify all definitions. Phenomena such as seasons, eclipses, phases of the Moon, etc. may be introduced too. Note that since the printed Human Orrery represent the correct elliptical orbits, it may be observed that Earth is closer to the Sun during winter in the northern hemisphere.

Human Orrery is best suited for the study of movements, including reference frames, notions of distance and duration related by speed, inertia, forces, and acceleration. There is no doubt that the difficulties associated with these notions are present at almost all levels in schools and for the general public too. Overall, cognitive schemata in the description of movements are very often adapted to the tasks of everyday life and thus lead to erroneous conceptions of more complex situations (Viennot, 1996). In the first place, the transition from one frame to another is difficult because it is not done in everyday life (Saltiel & Malgrange, 1980). In particular, the difference between heliocentric and geocentric landmarks has historically led to a paradigm shift. However, it is often misinterpreted as characterizing a fair model and a false model (Shen & Confrey, 2010). Secondly, the perception of speed difference is often understood assuming constant distances or constant durations. Taking simultaneously into account speed, distance and duration is problematic. This difficulty is clearly not solved by the usual teaching based on formulas or graphs (Trowbridge & McDermott, 1980; Thompson, 1994; Trudel & Métioui, 2011;). Finally, the concept of time is fundamental in an enacted framework as described here. The word "time" hides very different notions, whether biological, psychological, physical or measurement. We use the terms of Coquide & Morge (2011) which describe time as "the transformation of space", and, referring to Bergson's work, indicate that "duration is considered as characteristic of a vital momentum, partially biological and evolutionary, and in a philosophy of continuity ". We understand then the essential difficulty of an adequate understanding of time in the study of a movement at all levels of study, including university.

The next step is to investigate the causes of speed variation, or acceleration. The movements of bodies in the Solar System and in our Galaxy are governed by one and the same law (for large bodies with a diameter greater than 30-40 km): that of universal gravitation combined with the principle of inertia, in the framework of Newtonian mechanics. The accounting of Einsteinian mechanics is beyond our approach here. In this area, the prevalence of intuitive or incorrect notions of inertia into adulthood has often been studied (Viennot, 1996; Trudel & Métioui, 2011). It has been noticed that difficulties arise when attention is directed to the trajectory of the moving object (circular in particular). In these cases, the false link between the force on the object and its speed is difficult to challenge in the mind of the learner.

### 1.3. Enation Theory as a Theoretical Background

The use of a Human Orrery in education is based on the assumption that bodily perceptions help the learning of abstract concepts. The conversion of learning space into performance space means that embodiment becomes a vehicle for interpretation (Lindgren & Johnson-Glenberg, 2013) and movement a medium for choreographing thought (Lapaire, 2017). New connections are set up between mental activity and kinetic activity; sense-making and sensory-motor experience. Such an assumption is based on the cognition theory of enaction (Varela, 1991) that is already well known and widely used in Science and Mathematics Education, e.g. Johnson-Glenberg et al. (2016). In recent years, research on learning and education has been increasingly influenced by theories of embodied cognition. Several embodiment-based interventions have been empirically investigated, including gesturing, interactive digital media, and bodily activity in general. The setting-up of our designs bring us to the highest level of embodiment as defined by Johnson-Glenberg et al. (2014) as it is using integrated forms of embodied learning and a high level of bodily engagement. Enaction theory considers the interaction between the body and the outside world as the foundation of cognition, what Lapaire (2017) calls "Kineflexion" and Radford (2014) "a sensuous cognition". Abstract concepts must then be embodied and experienced in sensory experiences to be fully integrated (Glenberg, 2015); the environment becomes then an actor of reflection under the condition that it favours a fine perception (Varela, 1991). This distinction between fine perception and coarse perception proposed by Varela brings us closer to the conclusions of didactics.

To be in line with the enaction theory, as shortly described above, the use of the Human Orrery has to provide students with an adapted or congruent sensory pathway to the abstract notions to be learned (Segal, 2011). We refer the reader to Rollinde & Decamp (2019), which provides a "dictionary of the Human Orrery" that describes in detail how gestures and movements on the Orrery are associated with abstract concepts of kinematics.

### 2. Medialisation of Occultation

We propose in this section an illustration of the Human Orrery adapted to the topic of occultation. The first stages of this sequence, up to the observation of one planet that pass another one, has been tested in September 2019 with French 12-year-old pupils of the junior high school Pailleron located in Paris. Figures in the text are extracted from those sequences. The computational part of the sequence, namely the prediction of periodicity of occultation was tested a few years ago with 1st-year University students in France.

The sequence is based on the modelling principles (Halloun & Hestenes, 1985; Sensevy et al., 2008) that require regular connections by the learners between the experimental world and the world of theories and models. A model allows to represent salient characteristics of experimental objects; communicate about specificities of the objects; unify different classes of objects. The connection between the two worlds is made through different activities: (1) Within the experimental world, learner must use well-defined words, make connections, compare and classify objects and actions. (2) Through such activities, a model may emerge with specificities and general rules. (3) Those rules are used to make predictions of specific observations that will be tested in the experimental world. Failure of those predictions will force to come back to actions (1) and (2). (4) Manipulations within the model such as enrichment of the model, to better define concepts, to infer new predictions, to propose an application of the model for new experimental objects.

The Human Orrery is a model of the Solar System. Learners recognize the Solar System primarily by the presence of the Sun at the centre and occasionally through the identification of orbits. The model (representation of real objects) is thus already present with some specific features only (e.g. distances and not sizes). The experimental world cannot be accessed directly, neither duration of movements nor distance between objects. This ability to enact objects (here, planets) with our body and at our human scale is natural to human beings (Lapaire, 2019; McNeill, 1992). Modelling process must then be understood from an enacted experimental world to abstract models.

### 2.1. Dance of the Planets

The activity starts with a review of the planets, inquiring about "missing" ones on the Human Orrery. In the middle school where the sequence was tested, the Human Orrery includes only inner planets and one comet. Students easily spot the Sun. The names of planets emerge collectively. Then, some individuals in the group usually search for the moon (whatever the age of the group). They start by choosing one dot that corresponds to the comet Ceres and that is near one of the dot on Earth orbit. Some individuals may also choose Ceres or even Mercury whose image (when printed on the map) looks a bit like the moon. Discussing about the distance between Earth and the moon discards those possibilities: the moon would be under the dot of Earth and is thus not represented on the Human Orrery. This also allows the teacher to remark that the sizes of each object are not at scale. Going further, and following the position of moon orbiting around the Sun would prove that the moon's trajectory is at the first order of approximation a circle around the Sun.

The positions of the planets at the date of the day are then provided to the learners or obtained through a specific software (e.g. Stellarium©, Figure 2, left). Learners will then have to reproduce those positions on the Human Orrery (Figure 2, right). To do so, they need to search for specific geometrical configurations. Here, they reproduce the observed alignment of Earth (rightmost in the figure, blue dot), Sun and Mars (leftmost on the Figure, red dots). Some learners explicitly explained to the group that "Earth and Mars are opposite" (omitting the word Sun, that is implicit).

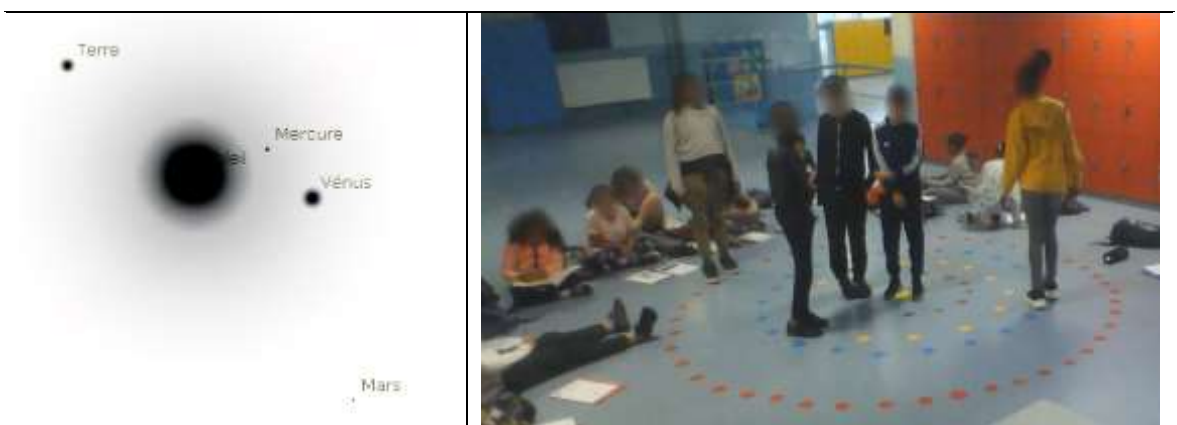

**Figure 2. Initial positions of the inner planets at the date of the activity as seen on ©Stellarium (left panel; in French) and reproduced by students (middle panel, from left to right: Mars, Venus, Mercury, Sun and Earth).**

The choreography may then be started. The teacher claps regularly, and each planet makes one step between two claps. All actor-planets reach the next point along its orbit at each single step, except

Mercury that jump every second point. At least one individual (usually Earth) tried to orbit around itself. This is obviously too difficult to be done for a long time. It must be explicated by the teacher that this specific feature is not conserved in the Human Orrery model.

Without explicit instruction by the teacher, each actor may start in a different direction. There is indeed no specific a-priori reason for all planets to orbit in the same direction. This may be an opportunity to discuss the global dynamic of a gravitational system (if appropriate given the age of the learners, and the duration of the activity) that ends up with all planets orbiting in the same direction.

Once all those aspects are fixed, all actor-planet walk with correct relative velocity and one may observe the evolution of their respective position with time. One classical prediction, not detailed here, is the evolution of the planets that may be observed from Earth in the nocturnal sky.

### 2.2. When Planets Cross Each Other

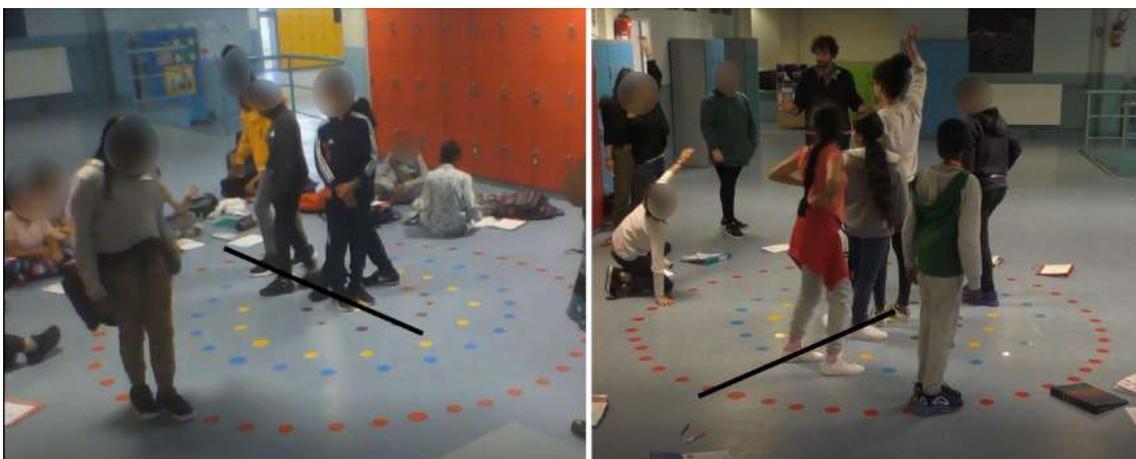

**Figure 3.Left: Venus (yellow dots) overtakes Earth (blue dots). Right: Mercury (brown dots, uneasy to distinguish here) overtakes Earth. The black line highlights the alignment of the two planets with the Sun.**

In many successive attempts, five learners play the role of the inner planet and the Sun while others observe. Different actors are selected each time so that all may experiment with the dance of the planets. A single instruction is given to both observers and actors: "raise your hand as soon as one planet pass by another one. Then, the actors must stop, and we will observe the respective position of the planets at that instant." Figure 3 provides the reader with two examples of such events, involving Mercury and Earth (right), and Venus and Earth (left). The definition of overtaking was quite obvious for all. Yet, we may notice that some learners raised their hand as soon as they observed that one planet was going faster than another one.

The gamification of the activity was obvious. Consequently, almost all individuals tried to spot overtaking, even those who were not interested in the activity at first. For a few attempts, the teacher lets the learners spot any planets. Then, the teacher asked to raise their hand only if Mars was involved in an overtaking. It then appears that Mars is never the one that pass by. The connection with the different speed was immediately perceived by most learners who almost shouted "but of course, Mars is travelling so slowly!". Next, the teacher asked to raise their hands only if Mercury was involved. Again, the connection with the speed of Mercury appears to be obvious for learners.

At each stop of the choreography, the teacher asked if observers have noticed something special in the relative position of planets. The answer was not obvious here. In one of the group, the Actor-Sun played a very important role here, revealing to others that s(he) was always in line with the two planets. In other groups, the teacher had to highlight the relative position. In all cases, the concept of alignment emerged at some point from the learners themselves.

The activity for middle school students stop here, with the written conclusion that (1) planets have different speed, (2) when one planet pass by another one, they are aligned with the Sun, (3) Mercury is the fastest planet and overtakes all planets while Mars is the slowest of all inner planet and never overtakes any inner planets. The activity was then illustrated by a photo of the transit of Venus.

**2.3. Prediction**

Playing the choreography during several terrestrial years, learners may directly observe and measure the periodicity of, e.g., the transit of Venus. This period is found to be 575 days… Note that many successive transits should be observed to check the actual periodicity of the transit. It should also become clear that the transit does not occur at the same position every 575 days. Students at University level were asked to create a simple iterative code that computes the position of Earth and Mars every 16 terrestrial days. They also compute the position that Venus should have along its orbit to create an occultation for every single position of Earth along its own orbit (the "occultation position", intersection of Venus orbit with the segment that connects the position of Earth to the Sun). They could then make a graph (Figure 4) showing the distance between the actual position of Venus to the occultation position related to the actual position of Earth every 16 days. The occultation is predicted every time this distance equals zero. Thus, the periodicity of the occultation is made obvious.

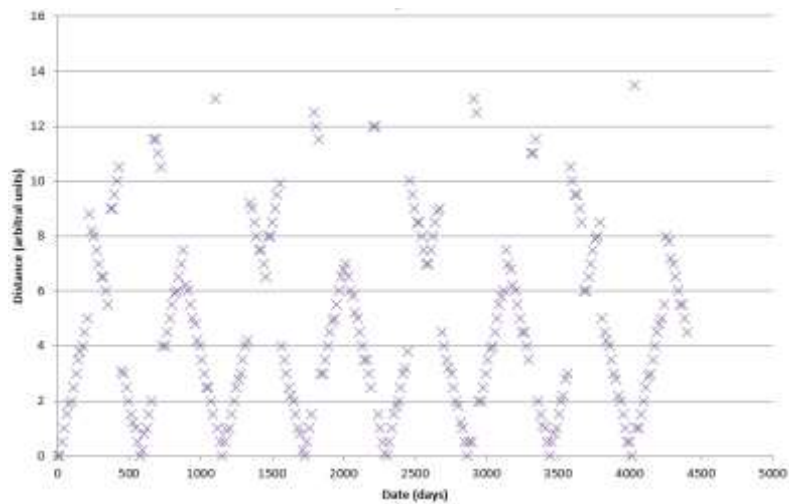

**Figure 4. Evolution of the distance to the occultation point for Venus and Earth with time (see text for details).**

Doing either one of the methods, one obtains a period of 575 days ±3 (this uncertainty is obtained with the numerical method). As explained in Section 2.1, orbital periods in the Human Orrery have different offset compared to the real ones for each planet. Then, after many turns, the actors are in advance or late along the orbits as compared to the real trajectory. This may be taken into account in both the choreography or the numerical code, and yields to a periodicity of 586 days ±3.

Those predictions may be compared to the theoretical values obtained by comparison of orbital periods. One may first wonder when Earth and Mars get back to the same positions simultaneously. This corresponds to 8 terrestrial years, 5 times the period we obtained. Yet, transit occur at different positions. The correct theoretical period is obtained using: $1/T = (1/T_{venus} - 1/T_{Earth})$, which corresponds to a periodicity of 584 days, in agreement with our prediction using the Human Orrery; However, observations of the real phenomena reveal that the transit of Venus occurs very rarely, with "pairs" of transit separated by 8 years with a periodicity of about 120 years.

This huge discrepancy forces the learners to question the validity of the model used for this activity. It appears that the Human Orrery is in 2D, while Venus orbital plane is not aligned with Earth orbital plane. Thus, the transit may occur only in two specific positions where the two planes intersect. This explains why the prediction of the model was wrong.

## 3. Conclusion and Perspectives

The perspective that we adopt in this work is an interdisciplinary approach that incorporates an embodied dimension and draws on real-world modelling. The use of the Human Orrery enables students to enact or "experience" scientific concepts and the dynamics of their properties by evolving in an adapted environment. In this contribution, we have detailed the use of a Human Orrery to help understand the occultation phenomena.

The current dynamic of the project is to motivate more classes from different levels and teachers from subjects other than mathematics and physics such as art, technology, music, languages (native or foreign), physical education and sport. In 2018-2019, more schools have decided to join the project in France: two primary schools in the context of interdisciplinary projects and one secondary school through a project involving science and technology. At the European level, a consortium is emerging involving researchers in mathematics and science didactics. Our next objective is then to design sequences that fit into the STEAM approach and that will be used as experimental situations for the theoretical framework of enaction, complemented with the framework of Activity Theory in the context of science education (Abboud et al. 2018). Our focus will be as much on the didactical learning as on the motivation and awareness of the learners during the activity.

## 4. Acknowledgments

The authors express their special thanks to the teachers of the junior high school Pailleron, who have always supported the implementation of the Human Orrery within their project « Des Pommiers sur Mars » (apple trees on Mars): M. Hauss, Mme Bessonies, Mme Bouchez Pommier, Mme Nardou, Mme Samali, M Saussez, and Arlette.